\documentclass[12pt]{article}

\usepackage{cite}
\usepackage{epsfig}
\usepackage{amsmath}
\usepackage{array}
\usepackage{bbold}

\def\mathswitch#1{\relax\ifmmode#1\else$#1$\fi}
\def\mathswitchr#1{\relax\ifmmode{\mathrm{#1}}\else$\mathrm{#1}$\fi}
\newcommand{\PW}{\mathswitchr W}

\newcommand{\scrs}{{}}
\newcommand{\sw}{\mathswitch {s_{\scrs\PW}}}
\newcommand{\cw}{\mathswitch {c_{\scrs\PW}}}

\newcommand{\tev}{\,\, \mathrm{TeV}}
\newcommand{\gev}{\,\, \mathrm{GeV}}

\newcommand{\SLASH}[2]{\makebox[#2ex][l]{$#1$}/}

\newcommand{\Eslash}{\SLASH{E}{.3}\,}
\newcommand{\RR}{{\rm R}}
\newcommand{\LL}{{\rm L}}
\newcommand{\HH}{{\rm H}}

\newcommand{\gesim}{\,{_{\textstyle
>}\atop^{\textstyle\sim}}\,}

\newcommand{\eps}{\epsilon}
\newcommand{\tr}{{\rm tr}}
\newcommand{\tp}{X}
\newcommand{\tpp}{T}
\newcommand{\tppp}{X- and T}

\newcommand{\mycaption}[1]{\caption{\sl #1}}

\begin{document}
\thispagestyle{empty}

\def\thefootnote{\fnsymbol{footnote}}

\begin{flushright}
ZU-TH 10/09
\end{flushright}

\vspace{1cm}

\begin{center}

{\Large\sc {\bf A Little Higgs Model with Exact Dark Matter Parity}}
\\[3.5em]
{\large\sc
A.~Freitas$^1$, P.~Schwaller$^2$,
D.~Wyler$^2$
}

\vspace*{1cm}

{\sl $^1$
Department of Physics \& Astronomy, University of Pittsburgh,\\
3941 O'Hara St, Pittsburgh, PA 15260, USA
}
\\[1em]
{\sl $^2$
Institut f\"ur Theoretische Physik,
        Universit\"at Z\"urich, \\ Winterthurerstrasse 190, CH-8057
        Z\"urich, Switzerland
}

\end{center}

\vspace*{2.5cm}

\begin{abstract}
Based on a recent idea by Krohn and Yavin, we construct a little Higgs model
with an internal parity that is not broken by anomalous Wess-Zumino-Witten
terms. The model is a modification of the ``minimal moose'' models by
Arkani-Hamed et al. and Cheng and Low.  The new parity prevents large
corrections to oblique electroweak parameters and leads to a viable dark matter
candidate. It is shown how the complete Standard Model particle content,
including  quarks and leptons together with their Yukawa couplings, can be
implemented. Successful electroweak symmetry breaking and consistency with
electroweak precision constraints is achieved for natural parameters choices. A
rich spectrum of new particles is predicted at the TeV scale, some of which have
sizable production cross sections and striking decay signatures at the LHC.
\end{abstract}

\def\thefootnote{\arabic{footnote}}
\setcounter{page}{0}
\setcounter{footnote}{0}

\newpage


\section{Introduction}

Little Higgs models are effective non-supersymmetric theories with a natural
cutoff scale at about 10~TeV, where the Higgs scalar is a pseudo-Goldstone boson
of a global symmetry, which is spontaneously broken at a scale $f\sim1$~TeV. 
The symmetry breaking pattern protects the Higgs mass from quadratically
divergent one-loop corrections, which are cancelled by new gauge bosons and
fermions with masses near $f$. Therefore the hierarchy of scales can be realized
without fine-tuning the parameters in the Higgs potential. A simple
implementation of this mechanism is given by the ``minimal moose'' model of
Ref.~\cite{minimalmoose}. This model has two copies of the Standard Model (SM)
gauge group, which are broken to the diagonal group at the scale $f$,
reminiscent of chiral symmetry breaking in QCD.

However, tree-level mixing between the gauge bosons introduces large corrections
to the oblique electroweak parameters for $f\sim1$~TeV, unless the gauge
couplings of the two gauge sectors are almost equal~\cite{Kilic:2003mq}. This
equality of couplings can be explained by a discrete symmetry called \tpp-parity
\cite{tpar, LHT}, under which the SM fields are \tpp-even and the new TeV-scale
particles are odd\footnote{A different discrete symmetry, which
does not lead to a complete doubling of the SM particle content, has been
proposed in Ref.~\cite{ArkaniHamed:2002pa,Chang:2003un}.}. As a result, all tree-level interactions between \tpp-even
and \tpp-odd particles are forbidden, so that corrections to the electroweak
precision observables occur only at one-loop level and thus are sufficiently
small to allow values of $f$ of 1~TeV and below. Furthermore, the lightest
\tpp-odd particle is stable and, if neutral, can be a good dark matter
candidate.

Often it is assumed that the new physics entering near the scale of 10~TeV
are some strong dynamics similar to technicolor theories\footnote{An alternative
approach involving a weakly coupled symmetry breaking sector
can be found in Ref.~\cite{Csaki}.}. In this case, however,
the fundamental theory can induce a Wess-Zumino-Witten (WZW) term \cite{WZW},
which is \tpp-odd \cite{hillsq} if \tpp-parity is implemented as in
Ref.~\cite{LHT}. The breaking of \tpp-parity by the WZW term,
though suppressed by the large symmetry breaking scale, rules out the lightest
\tpp-odd particle as a dark matter candidate, since this particle would decay
promptly into gauge bosons \cite{lhtwzw}.
On the other hand, it was recently shown that a different construction of
the parity in moose models leads to a parity-even WZW term \cite{Krohn:2008ye}.
The authors present a simple toy model that shows the relevant features.

In this article we adopt the idea of Ref.~\cite{Krohn:2008ye} for the
``minimal moose'' model in order to construct a fully realistic model
which reproduces the Standard Model as a low-energy theory, admits electroweak
symmetry breaking (EWSB), is consistent with
electroweak precision constraints, and has a viable dark matter candidate.
In the following section, the model and the implementation of the new 
\tp-parity is
described explicitly. In section~\ref{sc:masses} the physical mass spectrum of
the model is analyzed,
and it is shown that successful electroweak symmetry breaking can be achieved.
Finally, section~\ref{sc:pheno} discusses
electroweak precision constraints and gives a brief overview of the collider
phenomenology, before the conclusions are presented in section~\ref{sc:concl}.


\section{The model}
\label{sc:model}

The model is based on a large ${\rm SU(3)^8 = [SU(3)_\LL\times SU(3)_\RR]^{4}}$ global symmetry group that is spontaneously broken to the diagonal vector group ${\rm SU(3)_{\rm V}^4}$ at a scale $f$, giving rise to four sets of ${\rm SU}(3)$ valued nonlinear sigma model fields
\begin{equation}
X_i = e^{2ix_i/f}, \qquad i=1,\dots,4.
\end{equation}
Under the global symmetry group they transform as $X_{1,3} \to L_{1,3} X_{1,3} R_{1,3}^\dagger$ and $X_{2,4} \to R_{2,4} X_{2,4} L_{2,4}^\dagger$. The axial components of the global symmetries shift the Goldstone fields, $x_i \rightarrow x_i + \epsilon_i$, thereby forbidding any nonderivative couplings for the Goldstone fields. In particular, as long as these symmetries are not explicitly broken, a mass term can't be generated for the Goldstone fields at any loop order. 

Adding gauge and Yukawa interactions will in general break some of the global symmetries and therefore generate ${\cal O}(f)$ mass terms for the corresponding Goldstone bosons. The idea of collective symmetry breaking is to implement the required interactions in such a way that each interaction respects parts of the global symmetry and therefore keeps the corresponding Goldstone bosons massless. Only the simultaneous presence of different symmetry breaking interactions can then generate a mass for those Goldstone bosons. Since appropriate diagrams only appear at the two-loop level, the generated masses are suppressed by an additional loop factor and can be significantly below the scale $f$. 

Our goal is to have at least one light electroweak doublet that we can identify with the SM Higgs boson. Under the SM gauge interactions, the  Goldstone fields $x_i$ decompose as follows

\begin{align}
        x_{i} = 
        \begin{pmatrix}
                \phi_{i}+ \eta_{i}/\sqrt{12} & h_{i}/2 \\ h_{i}^{\dagger}/2 & -
                \eta_{i} / \sqrt{3}
        \end{pmatrix},
\end{align}
where $\phi_i=\phi_i^a\sigma^2/2$ are triplets under the SU(2) gauge group,
$h_i$ are complex doublets, and $\eta_i$ are real singlets. We further demand that the physical Higgs boson is even under the dark matter parity that acts as $x_1 \leftrightarrow x_2$ and $x_3 \leftrightarrow x_4$ on the Goldstone fields.
This leaves us with two candidates for the SM Higgs doublets, 
\begin{align}
    h_a &\equiv \tfrac{1}{\sqrt{2}} (h_3 + h_4),  &
    h_b &\equiv \tfrac{1}{\sqrt{2}} (h_1 + h_2). \label{higgsb}
\end{align}
The physical Higgs field will later be identified as $h_a$ and is protected by the global symmetries ${\rm SU(3)_{\rm L,a} = SU(3)_{\rm L,3} \times SU(3)_{\rm L,4}/SU(3)_{DL}}$ and ${\rm SU(3)_{\rm R,a} = SU(3)_{\rm R,3} \times SU(3)_{\rm R,4}/SU(3)_{DR}}$,\linebreak where ${\rm SU(3)_{Di}}$ denotes the diagonal subgroups of these product groups.  
As long as no single interaction breaks both  ${\rm SU}(3)_{\rm L,a}$ and ${\rm SU}(3)_{\rm R,a}$ at the same time, the mass of the Higgs will be sufficiently small. 

For models based on the  symmetry structure used here, possibilities to introduce interactions that preserve enough global symmetries are discussed in \cite{minimalmoose}. We found that we could adopt their rules to introduce scalar self-interactions as well as gauge interactions, but that some modifications are required in the Yukawa sector in order to maintain the parity symmetry. In particular partners for the standard model fermions must be introduced so that the dark matter parity can be implemented in a linear way. 
%

\subsection{Scalar and gauge sector}

\begin{figure}
\begin{center}
\ \hspace{15mm}
\psfig{figure=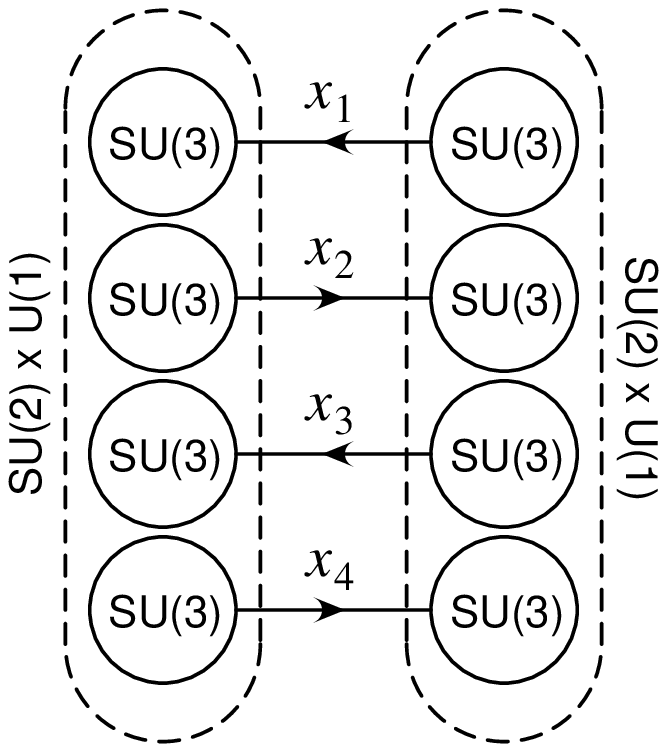, width=6.5cm}%
\end{center}
\vspace{-1em}
\mycaption{Illustration of the global and gauge symmetry structure of the model.
\label{moosediag}} 
\end{figure}

The global symmetry structure of the model is is depicted in Fig. \ref{moosediag}. 
On each site, a ${\rm SU}(2)\times {\rm U}(1)$ subgroup is gauged, with equal strength for
both sites. The gauge group generators are given by
\begin{align}
Q_{\LL,\RR}^a &= \begin{pmatrix} \sigma^a/2 & 0 \\ 0 & 0 \end{pmatrix}, 
&
Y_{\LL,\RR} &= \frac{1}{\sqrt{12}}\begin{pmatrix} \mathbb{1} & 0 \\ 0 & -2 \end{pmatrix},
\end{align}
written in terms of $2\times2$ and $1\times1$ blocks. Here $\sigma^a$ denote the
Pauli matrices. The kinetic term of the sigma fields reads
\begin{align}
\label{LG}
{\cal L}_{\rm G} &= \frac{f^2}{4} \sum_{i=1}^4 \tr [(D_\mu X_i)(D^\mu
X_i)^\dagger], & \text{with}&&
&D_{\mu} X_{1,3} = \partial_{\mu}X_{1,3} - i A_{\LL\mu} X_{1,3}
					+ i X_{1,3}A_{\RR\mu}, 
\\[-2ex]
&&&& \nonumber
&D_{\mu} X_{2,4} = \partial_{\mu}X_{2,4} - i A_{\RR\mu} X_{2,4}
					+ i X_{2,4}A_{\LL\mu},
\\[1ex] && \text{and} &&
&A_{\LL\mu} \equiv g_\LL W_{\LL\mu}^a Q_\LL^a + g'_\LL y_{\LL X} \, B_{\LL\mu} Y_\LL,
\\ &&&&
&A_{\RR\mu} \equiv g_\RR W_{\RR\mu}^a Q_\RR^a + g'_\RR y_{\RR X} \, B_{\RR\mu} Y_\RR,
\end{align}
where the gauge couplings at the two sites are chosen to be equal,
$g_\LL=g_\RR=\sqrt{2}g$ and $g'_\LL=g'_\RR=\sqrt{2}g'$, and $g,g'$ are the SM
gauge couplings. Furthermore, $y_{\LL X,\RR X}$ denote the U(1) charges of the
fields $X_i$. The choice $y_{\LL X}=y_{\RR X}=1/\sqrt{3}$ ensures the correct
values for the Higgs doublet hypercharge and Weinberg angle.
Note that the definition \eqref{LG} of the covariant
derivatives corresponds to assigning  opposite directions for the link fields
1,3 and 2,4, which is important for the definition of the \tp-parity below.

Each gauge interaction separately only break either ${\rm SU(3)_{L,a}}$ or ${\rm SU(3)_{R,a}}$ and therefore respects collective symmetry breaking. Actually since the gauge interactions are either on the left or on the right side of the moose diagram, no large mass is generated for any of the Goldstone fields from these interactions. 

The kinetic term \eqref{LG} has a $\mathbb{Z}_2$ symmetry, called \tp-parity,
defined by
\begin{align}
\label{tpar}
\text{\tp-parity:} \qquad
	A_\LL\leftrightarrow A_\RR\, , 
\qquad  X_{1}\leftrightarrow X_{2}\, , 
\qquad  X_{3}\leftrightarrow X_{4}\, .
\end{align}
This definition is a straightforward generalization of
the parity of the two-link model in Ref.~\cite{Krohn:2008ye}. Under this parity,
the WZW terms \cite{WZW} for the four link fields transform as
\begin{align} 
\Gamma_{\rm WZW}(x_1,A_\LL,A_\RR) &\leftrightarrow
  \Gamma_{\rm WZW}(x_2,A_\RR,A_\LL), &
\Gamma_{\rm WZW}(x_3,A_\LL,A_\RR) &\leftrightarrow
  \Gamma_{\rm WZW}(x_4,A_\RR,A_\LL),
\end{align}
so that the combined term 
\begin{equation}
{\cal L}_{\rm WZW} = \Gamma_{\rm WZW}(x_1,A_\LL,A_\RR) +
		     \Gamma_{\rm WZW}(x_2,A_\RR,A_\LL) +
		     \Gamma_{\rm WZW}(x_3,A_\LL,A_\RR) +
		     \Gamma_{\rm WZW}(x_4,A_\RR,A_\LL)
\end{equation} 
remains invariant.
As a result, \tp-parity is an exact symmetry of the model and the lightest \tp-odd
particle is stable.

In addition to the \tp-parity in eq.~\eqref{tpar} a second $\mathbb{Z}_2$ symmetry,
called \tpp-parity, is imposed, under which
\begin{align}
\label{tppar}
\text{\tpp-parity:} \qquad\qquad
	A_\LL\leftrightarrow A_\RR\, , 
\qquad  X_{i}\to \Omega X_{i}^\dagger \Omega \, , 
\end{align}
where $\Omega \equiv {\rm diag}(1,1,-1)$. Our \tpp-parity is identical to the
original version in Ref.~\cite{LHT}, and it ensures that the triplet and singlet
scalar do not receive any vacuum expectation values. In our implementation,
\tpp-parity is respected by the model at the classical level, 
but broken by ${\cal L}_{\rm WZW}$. However, since the stability of the dark
matter candidate is already guaranteed by \tp-parity \eqref{tpar}, this does not
lead to any problems.

In the gauge sector, the \tp-odd linear combinations of gauge bosons,
\begin{equation}
W_\HH^a = \tfrac{1}{\sqrt{2}}(W_\LL^a-W_\RR^a), \qquad
B_\HH = \tfrac{1}{\sqrt{2}}(B_\LL-B_\RR),
\end{equation}
acquire masses of order $f$ from the kinetic term \eqref{LG}, while the
\tp-even combinations
\begin{equation}
W^a = \tfrac{1}{\sqrt{2}}(W_\LL^a+W_\RR^a), \qquad
B = \tfrac{1}{\sqrt{2}}(B_\LL+B_\RR),
\end{equation}
remain massless before EWSB and are identified with the SM gauge bosons.
The scalar fields form the following \tp-even and \tp-odd combinations:
\begin{align}
	w &= \frac{1}{2}(x_{1}-x_{2}+x_{3}-x_{4}) &  x &=
	\frac{1}{2}(-x_{1}+x_{2}+x_{3}-x_{4})  && \text{(\tp-odd)}, \\
	y &= \frac{1}{2}(-x_{1}-x_{2}+x_{3}+x_{4}) &  z &=\frac{1}{2}(
	x_{1}+x_{2}+x_{3}+x_{4}) && \text{(\tp-even)}.
\end{align}
The triplet $\phi_w$ and the singlet $\eta_w$ are eaten to form the longitudinal
components of $W_\HH^a$ and $B_\HH$.

A large Higgs quartic coupling, required for electroweak symmetry breaking, is
generated by the following \tp-invariant plaquette operator:
\begin{align}
	{\cal L}_{\rm P} = \frac{\kappa}{8}  f^{4} \, \tr\left[
X_{1}X_{3}^{\dagger}X_{2}^{\dagger}X_{4} +
X_{2}X_4^{\dagger}X_{1}^{\dagger}X_{3} \right] + \text{h.c.}
\label{plaquette}
\end{align}
This operator contains an explicit ${\cal O}(f)$ mass term for the scalar fields in $x$, but preserves enough global symmetries so it does not generate large masses for any other Goldstone bosons at the one loop level, in particular not for $h_a$, $h_b$.

Successful electroweak symmetry also requires the introduction of a second
plaquette term \cite{minimalmoose}, which breaks a different subset of the
global symmetry:
\begin{align}
    {\cal L}'_{\rm P} = \frac{\epsilon}{8} f^4 \tr \left( T_8 X_1 X_3^\dagger X_2^\dagger X_4 + T_8 X_2X_4^\dagger X_1^\dagger X_3 +X_1 X_3^\dagger T_8 X_2^\dagger X_4 +  X_2X_4^\dagger T_8 X_1^\dagger X_3 \right) + \text{h.c.}
    \label{epsplaq}
\end{align}
where $T_8 = {\rm diag}(1,1,-2)/\sqrt{12}$, and $\epsilon$ is a complex 
constant. 
As explained in Ref.~\cite{minimalmoose}, eq.~\eqref{epsplaq}
can be generated radiatively by two-loop diagrams involving the top quark, and
therefore it is natural to assume that $|\epsilon| \sim |\kappa|/10$.
We can assume $\epsilon$ to be purely imaginary, since the real part only
gives small corrections to the scalar potential.


\subsection{Fermion sector}

For the construction of the kinetic and Yukawa terms of the fermions, several
conditions need to be considered. First, one has to make sure that these terms
do not break too many of the global symmetries, so that the mass of the little
Higgs doublet remains protected from quadratic corrections. Secondly, the
minimal construction using only \tp-even fermions \cite{LHT} leads to unsuppressed
four-fermion operators at one-loop level, thus forcing the scale $f$ be about
10~TeV or larger~\cite{Low:2004xc}. The second problem can be solved by 
introducing ``mirror'' fermions \cite{Low:2004xc}, {\it i.$\,$e.} two sets of
fermions that are partners under \tp-parity. Our implementation closely
resembles the setup in the appendix of Ref.~\cite{Cheng:2005as}.

For each SM flavor two doublets of left-handed fermions
are introduced, located at the two sites of the moose diagram.
With the exception of the top quark,
they are embedded into incomplete representations of SU(3) as follows
\begin{equation}
  Q_{\rm a}=(d_{\rm a}, u_{\rm a}, 0)^\top, \qquad\qquad  
  Q_{\rm b}=(d_{\rm b}, u_{\rm b}, 0)^\top.
  \label{lightferm}
\end{equation}
Under the global ${\rm SU(3)_\LL\times SU(3)_\RR}$ group they transform as $Q_{\rm a}
\to L_i Q_{\rm a}$ and $Q_{\rm b} \to R_i Q_{\rm b}$, while \tp- and \tpp-parity
interchange the two fields, $Q_{\rm a} \leftrightarrow Q_{\rm b}$.

Since \eqref{lightferm} are incomplete multiplets, their interaction
terms break the global symmetries that protect the Higgs mass 
and lead to quadratically divergent
contributions from one-loop diagrams involving the Yukawa couplings.
For the first two generations this is not a problem since the Yukawa couplings
are very small, but for the third generation we need to introduce complete
multiplets
\begin{equation}
  Q_{\rm 3a}=(d_{\rm 3a}, u_{\rm 3a}, U_{\rm a})^\top, \qquad\qquad  
  Q_{\rm 3b}=(d_{\rm 3b}, u_{\rm 3b}, U_{\rm b})^\top.
\end{equation}
Here the additional singlets $U_{\rm a,b}$ cancel the quadratically divergent
Higgs mass contributions induced by the large top Yukawa coupling.

The \tppp-invariant fermion kinetic terms have the standard form
\begin{align}
{\cal L}_{\rm F} &= 
 i\overline{Q}_{\rm a} \bar{\sigma}^\mu D^{\rm a}_\mu Q_{\rm a} +
 i\overline{Q}_{\rm b} \bar{\sigma}^\mu D^{\rm b}_\mu Q_{\rm b}, 
& \text{with}&&
&D^{\rm a}_\mu = \partial_\mu + i g_\LL W_{\LL\mu}^a \,(Q_\LL^a)^\top
	- i g'_\LL \, y_{\LL Q} \, B_{\LL\mu}, \\
&&&&&
D^{\rm b}_\mu = \partial_\mu + i g_\RR W_{\RR\mu}^a (Q_\RR^a)^\top 
	- i g'_\RR y_{\RR Q} \, B_{\RR\mu}, \nonumber
\end{align}
where $\bar{\sigma}^\mu \equiv (1,-\vec{\sigma})$, and $y_{\LL Q}$ and $y_{\RR
Q}$ are diagonal matrices composed of the U(1) charges in Table~\ref{ferm}.
The SM fermions emerge from the \tp-even linear combination $Q =
\frac{1}{\sqrt{2}} (Q_{\rm a}+Q_{\rm b})$. To give mass to the \tp-odd
combination $Q_\HH = \frac{1}{\sqrt{2}} (Q_{\rm a}-Q_{\rm b})$, we need to
introduce conjugate Dirac partners
\begin{equation}
  Q_{\rm c}^c=(d_{\rm c}^c, u_{\rm c}^c, 0)^\top,
  \qquad\qquad
  Q_{\rm 3c}^c=(d_{\rm 3c}^c, u_{\rm 3c}^c, U_{\rm c}^c)^\top,
\end{equation}
Under ${\rm SU(3)_\LL\times SU(3)_\RR}$ they transform as $Q_{\rm c}^c
\to U_i Q_{\rm c}^c$, where $U_i$ ($i=1,\dots,4$) belongs to the unbroken diagonal
subgroup of $SU(3)_\LL\times SU(3)_\RR$ and is a non-linear function of $L_i$
and $R_i$. Furthermore, the effect of \tppp-parity is defined as
$Q_{\rm c}^c \to -\Omega Q_{\rm c}^c$. Then 
a \tppp-invariant mass term for the \tp-odd fermions is given by
\begin{equation}
  {\cal L}_{\rm M} = -\frac{\lambda_{c}}{\sqrt{2}} f \left( Q_{\rm a} \xi_{1} Q_{\rm c}^{c} 
  	- Q_{\rm b}\Omega \xi_{1}^{\dagger} Q_{\rm c}^{c} 
	- Q_{\rm b}\xi_{2}\Omega Q_{\rm c}^{c} 
	+ Q_{\rm a}\Omega \xi_{2}^{\dagger}\Omega Q_{\rm c}^{c}\right) 
	+ \mathrm{h.c.}\, , \label{yuk1}
\end{equation}
where $\xi_i = e^{ix_i/f}$. Under global $SU(3)_\LL\times SU(3)_\RR$ rotations
$\xi_i$ transforms as $\xi_i \to L_i \xi_i U_i^\dagger = U_i \xi_i R_i^\dagger$ for $i=1,3$ and analogous for $i=2,4$,
so that eq.~\eqref{yuk1} is evidently gauge invariant. 
In general, $\lambda_c$ is a $3\times3$ matrix in flavor space. Since it can
contribute to flavor-changing neutral currents (FCNCs) at one-loop level, it is
constrained by data on heavy-flavor decays and oscillations.  
Such effects are studied for example in \cite{Blanke:2006eb} for the case of
the littlest Higgs model with T-parity. 
For the analyses in section 3 and 4 we assume a flavor diagonal $\lambda_c$ for simplicity. 

Since the $Q_{\rm c}^c$ transform non-linearly, one must make use of the
$\xi_i$ fields to construct a gauge- and \tppp-invariant kinetic term.
Following the formalism of Callan, Coleman, Wess, and Zumino \cite{CCWZ}, 
it can be written as
\begin{align}
        {\cal L}_{\rm c} &= i\overline{Q}_{\rm c}^c \, \bar\sigma^{\mu} 
	\left ( \partial_{\mu} + \frac{1}{4} ( 
	  \xi_{1}^{\dagger} D_{\mu} \xi_{1} + 
	  \xi_{1} D_{\mu} \xi_{1}^{\dagger} +
	  \xi_{2}^{\dagger} D_{\mu} \xi_{2} + 
	  \xi_{2} D_{\mu} \xi_{2}^{\dagger}) 
	- i g' (y_{Q_{\rm c}}+\tfrac{1}{\sqrt{3}}Y_{\rm V}) B_{\mu} \right)
	Q_{\rm c}^c \, , 
\label{nonlin}
\intertext{where}
&\xi_{i}^{\dagger} D_{\mu} \xi_{i} =
	\xi_{i}^{\dagger} (\partial_\mu + igW^aQ^a_{\rm V} 
		+ igW^a_\HH Q^a_{\rm A} 
		+ ig'\tfrac{1}{\sqrt{3}} BY_{\rm V} 
		+ ig'\tfrac{1}{\sqrt{3}} B_\HH Y_{\rm A}) 
	\xi_{i}, \\
&\xi_{i} D_{\mu} \xi_{i}^{\dagger} =
	\xi_{i} (\partial_\mu + igW^aQ^a_{\rm V} 
		- igW^a_\HH Q^a_{\rm A} 
		+ ig'\tfrac{1}{\sqrt{3}} BY_{\rm V} 
		- ig'\tfrac{1}{\sqrt{3}} B_\HH Y_{\rm A}) 
	\xi_{i}^{\dagger}, 
\end{align}
and $Q^a_{\rm V}, Y_{\rm V}$ and $Q^a_{\rm A}, Y_{\rm A}$ are the unbroken and
broken gauge generators, respectively. Both equations (\ref{yuk1}) and (\ref{nonlin}) do not involve the $x_3$ and $x_4$ Goldstone fields and therefore do not break the global symmetries that protect the Higgs mass. They do however generate masses for some of the other Goldstone bosons that will be explicitly calculated in section \ref{sc:scalars}. 

Now Yukawa couplings can be constructed for the \tp-even massless combinations
of the fermions. For the up-type quarks of the first two
generations they read
\begin{equation}
{\cal L}_{\rm u} = -\lambda_{\rm u} f 
  Q_{\rm a} (X_3 + \Omega X_4^\dagger\Omega) 
  	\begin{pmatrix} 0 \\ 0 \\ u^c \end{pmatrix}
- \lambda_{\rm u} f Q_{\rm b} (\Omega X_3^\dagger\Omega + X_4) 
	\begin{pmatrix} 0 \\ 0 \\ u^c \end{pmatrix} + \mathrm{h.c.}
	\label{upyuk},
\end{equation}
where $u^c$ is are the right-handed quarks (one for each flavor), which are
\tppp-even. 
As already mentioned above, the presence of incomplete multiplets in the Yukawa
couplings leads to quadratically divergent contribution to the Higgs mass.
Therefore the top Yukawa coupling has a slightly different form \cite{Cheng:2005as}, 
\begin{equation}
{\cal L}_{\rm t} = -\lambda f 
  Q_{\rm 3a} (X_3 + \Omega X_4^\dagger\Omega) 
  	\begin{pmatrix} 0 \\ 0 \\ U_{\rm b}^c \end{pmatrix}
- \lambda f Q_{\rm 3b} (\Omega X_3^\dagger\Omega + X_4) 
	\begin{pmatrix} 0 \\ 0 \\ U_{\rm a}^c \end{pmatrix} + \mathrm{h.c.}
	\label{topyuk}.
\end{equation}
Here the two singlets $U_{\rm a}^c$ and $U_{\rm b}^c$ transform under
\tppp-parity as $U_{\rm a}^c \leftrightarrow U_{\rm b}^c$. Their
\tp-even combination $U_{\rm a}^c + U_{\rm b}^c$  emerges in the right-handed
top quark, while the \tp-odd combination $U_{\rm a}^c - U_{\rm b}^c$ forms the
right-handed partner of the \tp-odd  $U_{\rm a} - U_{\rm b}$. In addition there
are one more \tp-even and \tp-odd fermion in the top sector, which receive
masses from eq.~\eqref{yuk1}. This will be
explained in more detail in section~\ref{topm}. 

The use of complete multiplets $ Q_{\rm 3a}$, $Q_{\rm 3b}$ in (\ref{topyuk}) makes sure that each term preserves one of the global $\rm SU(3) $ symmetries that protect the Higgs mass. 

Finally, the down-type Yukawa couplings are given by
\begin{equation}
{\cal L}_{\rm d} = -\lambda_{\rm d} f 
  \widetilde{Q}_{\rm a} (X_3 + \Omega X_4^\dagger\Omega)^*
  	\begin{pmatrix} 0 \\ 0 \\ d^c \end{pmatrix}
- \lambda_{\rm d} f \widetilde{Q}_{\rm b} (\Omega X_3^\dagger\Omega +
	X_4)^* 
	\begin{pmatrix} 0 \\ 0 \\ d^c \end{pmatrix} + \mathrm{h.c.}
	\label{downyuk},
\end{equation}
where 
\begin{equation}
    \widetilde{Q}_{\rm a,b} = -2i T_2 Q_{\rm a,b} = 
    (-u_{\rm a,b}, d_{\rm a,b}, 0)^\top\,,
    \qquad
    \qquad
    T_2 = \begin{pmatrix} \sigma^2/2 & 0 \\ 0 & 0 \end{pmatrix}.
\end{equation}
The lepton Yukawa interactions are defined similarly. In contrast to the up-type
Yukawa couplings, the all three generations of  down-type fermions generate
quadratically divergent contributions to the Higgs doublet masses from
eq.~\eqref{downyuk},  which is permissible since the bottom Yukawa coupling is
much smaller than the top Yukawa coupling.
The kinetic term for the singlet conjugate fields $\psi^c \equiv u^c, d^c,
U_{\rm a}^c, U_{\rm b}^c$ simply reads 
\begin{equation}
    {\cal L}_{\rm R} 
= i\overline{\psi^c} \sigma^\mu (\partial_\mu - ig'y_{\psi^c}B_\mu)
  \psi^c
= i\overline{\psi^c} \sigma^\mu \bigl (\partial_\mu 
	- i\sqrt{2}g'(y_{\LL\psi^c}B_{\LL\mu}+y_{\RR\psi^c}B_{\RR\mu})
  \bigr )\psi^c,
\end{equation}
where $\sigma^\mu \equiv (1,\vec{\sigma})$ and
$y_{\psi^c} = 2y_{\LL\psi^c} = 2y_{\RR\psi^c}$ is the fermion hypercharge.

Table~\ref{ferm} summarizes the fermion contained in the model and their 
transformation properties.
\begin{table}
\renewcommand{\arraystretch}{1.2}
\begin{center}
\begin{tabular}{|c|c|c|c|c|c|c|}\hline 
 & ${\rm SU(2)_\LL}$ & ${\rm SU(2)_\RR}$ & ${\rm U(1)_\LL}$ & ${\rm U}(1)_\RR$ & \tp & \tpp \\ \hline \hline  
$q_{\rm a}$  & 2 & 1 & $\frac{1}{12}$ & $\frac{1}{12}$ & $q_{\rm b}$ & $q_{\rm b}$ \\ 
$ U_{\rm a}  $   & 1 & 1 & $\frac{7}{12}$ & $\frac{1}{12}$ & $U_{\rm b}$ & $U_{\rm b}$  \\
\hline 
$q_{\rm b}$   & 1 & 2 & $\frac{1}{12}$ & $\frac{1}{12}$ & $q_{\rm a}$ & $q_{\rm a}$ \\ 
$ U_{\rm b}  $   & 1 & 1 & $\frac{1}{12}$ & $\frac{7}{12}$ & $U_{\rm a}$ & $U_{\rm a}$ \\
\hline 
$Q^c_{\rm c}$   & \multicolumn{4}{c|}{nonlinear}  & $-\Omega Q_{\rm c}^c$ & $-\Omega Q_{\rm c}^c$ \\
\hline  
$ d^c  $   & 1 & 1 & $\frac{1}{6}$ & $\frac{1}{6}$ & $d^c$ & $d^c$ \\
$ u^c  $   & 1 & 1  & $-\frac{1}{3}$ & $-\frac{1}{3}$ & $u^c$ & $u^c$  \\ 
\hline 
$ U^c_{\rm a}  $   & 1 & 1 & $-\frac{7}{12}$ & $-\frac{1}{12}$ & $U_{\rm b}^c$ & $U_{\rm b}^c$ \\
$ U^c_{\rm b}  $   & 1 & 1  & $-\frac{1}{12}$ & $-\frac{7}{12}$ & $U_{\rm a}^c$ & $U_{\rm a}^c$  \\ 
\hline 
\end{tabular}
\mycaption{Quantum numbers of the fermion multiplets under the ${\rm [SU(2)\times
U(1)]^2}$ gauge symmetry, and their transformation properties under \tp\ and
\tpp.  The physical ${\rm U}(1)_{\rm Y}$ hypercharge is the sum of both
${\rm U}(1)_1+{\rm U}(1)_2$ charges. There is some freedom in the assignment of ${\rm U}(1)_1$ and
${\rm U}(1)_2$ charges to $U_{\rm a}^{(c)}$, $U_{\rm b}^{(c)}$. 
Here the conventions of \cite{Cheng:2005as} have been adapted.
\label{ferm}} 
\end{center}
\end{table}
Note that the model is non-renormalizable and considered to be a low-energy
effective theory of some fundamental dynamics associated with the UV cutoff
scale $\Lambda \sim 10 \, f \sim 10$~TeV. This UV completion could, but does not
need to, consist of some strongly coupled gauge interaction, which breaks the
global symmetry through the formation of a fermion condensate,
similar to technicolor.


\section{Mass spectrum}
\label{sc:masses}


\subsection{Top quark sector}
\label{topm}

Expanding the Yukawa couplings \eqref{yuk1} and \eqref{topyuk} in the top quark
sector in powers of $1/f$ yields 
\begin{align}
{\cal L}_{\rm t}=& -\sqrt{2} \lambda_c f (u_{\rm 3a} - u_{\rm 3b}) u_{\rm 3c}^c 
    		   -\sqrt{2} \lambda_c f (U_{\rm a} + U_{\rm b}) U^c_c 
		   -2 \lambda f \left( U_{\rm a} U^c_{\rm b} 
		   	+ U_{\rm b} U^c_{\rm a} \right) \nonumber\\ 
		 & -\lambda \left( q_{\rm 3a}(h_y+h_z) U^c_{\rm b} 
		 	+ q_{\rm 3b}(h_y+h_z) U^c_{\rm a} \right) \nonumber \\
                &+ {\scriptstyle \frac{1}{2\sqrt{2}}} \lambda_{c} \left[(q_{3a}+q_{3b}) (h_y-h_z) U^c_c + (U_{a}-U_{b}) (h_{y}^{\dagger}-h_{z}^{\dagger}) q_{c}^{c} \right]
		   +\dots + \mathrm{h.c.}\,,
		 \label{topyukraw}
\end{align}
where $q_{\rm 3a}=(d_{\rm 3a}, u_{\rm 3a})^\top, \; q_{\rm 3b}=(d_{\rm 3b},
u_{\rm 3b})^\top$, and the dots indicate ${\cal O}(f^{-1})$ terms and ${\cal
O}(f^{0})$ terms that do not involve Higgs doublets. With suitable phase
redefinitions of the fields, both $\lambda$ and $\lambda_c$ can be chosen to be
real\footnote{A relative factor $i$ between the second line of
\eqref{topyukraw} and \eqref{topyuk}
has been absorbed by this same procedure.}. Introducing the \tp-even
and -odd combinations
\begin{align}
U_\pm &\equiv \frac{1}{\sqrt{2}}(U_{\rm a} \pm U_{\rm b}), &
U_\pm^c &\equiv \frac{1}{\sqrt{2}}(U_{\rm a}^c \pm U_{\rm b}^c),  \\
q_{3\pm} &\equiv \frac{1}{\sqrt{2}}(q_{\rm 3a} \pm q_{\rm 3b}), &
u_{3\pm} &\equiv \frac{1}{\sqrt{2}}(u_{\rm 3a} \pm u_{\rm 3b}), 
\end{align}
one obtains
\begin{align}
    {\cal L}_{\rm t}  =& 
  -2 \lambda_c f u_{3-} u_{\rm 3c}^c - 2 \lambda_c f U_+ U^{\rm c}_c - 2
    \lambda f \left( U_+ U_+^c + U_-U_-^c \right) \nonumber \\
     &- \lambda 
    \left(q_{3+} (h_y+h_z) U_+^c + q_{3-} (h_y+h_z) U_-^c \right) + {\scriptstyle\frac{1}{2}}\lambda_c  q_{3+} (h_y-h_z) U^c_c + \mathrm{h.c.}\,
\label{topyukcooked}
\end{align}
Neglecting contributions of order $v^2/f^2$, the \tp-odd mass eigenstates in the
top sector, written in terms of left- and right-handed components, are
\begin{align} 
(T_\HH,T_\HH^c) &\equiv (u_{3-},u^c_{3c}), 
&
(T',T'^c) &\equiv (U_-,U_-^c), 
\end{align}
with masses $2\lambda_c f$ and $2 \lambda f$, respectively. In the \tp-even
top sector, the following Dirac fermions are formed:
\begin{align} 
(T,T^c) &\equiv \left( U_+, \; \frac{\lambda_c U_c^c + \lambda
  U_+^c}{\sqrt{\lambda^2 +\lambda_c^2}} \right), 
&
(t,t^c) &\equiv \left( u_{3+}, \; \frac{\lambda_c U^c_+ - \lambda U^c_c}{\sqrt{\lambda^2 +
  \lambda_c^2}} \right).
\end{align}
The $T$ obtains a mass $m_T=2 \sqrt{\lambda^2 + \lambda_c^2}
f$, while the SM-like top quark $t$ remains massless before EWSB and has a
Yukawa coupling given by
\begin{align}
    &-\lambda_t q_3 h t^c + \mathrm{h.c.}\,, & 
    \lambda_t &= \frac{\sqrt{2} \,\lambda \lambda_c}{\sqrt{\lambda^2 + \lambda_c^2}}\,.
    \label{tyuk}
\end{align}
Note that the \tp-odd top partner $T'$ is responsible for the cancellation of
the quadratically divergent contribution to the Higgs mass. Therefore the
\tp-even $T$ as well as the \tp-odd $T_\HH$ can be given masses of several TeV
by increasing $\lambda_c$, thus effectively decoupling them from the low-energy
theory.
The remaining fermion masses can be found in table \ref{partlist}.

Once electroweak symmetry is broken mixing of the top quark with the $T$ quark is reintroduced. The resulting mass matrix can be diagonalized by redefining the $t$ and $T$ quark as follows:
\begin{align}
    t & \rightarrow c_L t - s_L T, & T&\rightarrow c_L T + s_L t, \label{tmixL}\\
    t^c & \rightarrow c_R t^c - s_R T^c, & T^c &\rightarrow c_R T^c + s_R t^c,\label{tmixR} 
\end{align}
where $s_L\equiv\sin\alpha_L$, $c_L\equiv\cos\alpha_L$ are the sine and cosine 
of the left-handed mixing angle and similarly for $s_R$, $c_R$. To leading order
in an expansion in $(v/f)$, these mixing angles are given by
\begin{align}
    \sin\alpha_L &\approx \alpha_L = \frac{\lambda}{\lambda_c} \frac{m_t}{m_T} + {\cal O}\left(\frac{m_t^2}{m_T^2}\right), \label{sintexp}\\
    \sin\alpha_R &\approx \alpha_R = 0 + {\cal O}\left(\frac{m_t^2}{m_T^2}\right),
\end{align}
while the mass eigenvalues remain unperturbed at this order.


\subsection{Scalar masses}
\label{sc:scalars}

Since the non-linear sigma model breaks the complete symmetry down to its
diagonal vector group, the \tp-odd SU(2) and U(1) gauge bosons, which are
associated with the broken generators, become massive by eating the triplet
$\phi_w$ and singlet $\eta_w$ in the scalar $w$ multiplet, respectively. The
other scalars are pseudo-Goldstone bosons that receive masses from all
interactions that explicitly break some of the global symmetries. The only
tree-level mass terms for the scalars stem from the plaquette operators
\eqref{plaquette} and \eqref{epsplaq}, which lead to a mass $M_p^2 = 4\kappa 
f^2$ for all fields in the $x$ multiplet,  and additional ${\cal O}(\epsilon
f^2)$ contributions to all doublet fields. However, at one- and two-loop level,
scalar mass terms are generated from various other Lagrangian.

One-loop corrections from the mirror fermion mass term \eqref{yuk1} induce a
quadratically divergent mass for the linear combination $x_1+x_2 = -(y-z)$, of
order ${\cal O}[\lambda^2_c \Lambda^2/(16 \pi^2)] \sim {\cal O}(\lambda^2_c
f^2)$. Similarly, the top Yukawa coupling \eqref{topyuk} generates quadratically
divergent one-loop mass terms, of order ${\cal O}(\lambda^2 f^2)$  for the
doublets in $x_3-x_4 = x+w$ and singlets in $x_3+x_4 = y+z$. On the other hand,
the kinetic term \eqref{nonlin} leads to two-loop mass terms that have quartic
divergences \cite{LHT}. As a result, the scalar doublets in
$x_1^2+x_2^2=\frac{1}{2}(w-x)^2+\frac{1}{2}(y-z)^2$ pick up masses of order
${\cal O}[g^2/(16 \pi^2)^2 \times \Lambda^4/f^2] \sim {\cal O}(g^2 f^2)$.

The remaining doublet and triplet linear combinations $h_y+h_z$ and
$\phi_y+\phi_z$ are protected from quadratically divergent one-loop mass terms. 
However, all scalar fields obtain logarithmic one-loop contributions and
quadratically divergent two-loop contributions from the gauge kinetic terms and
the plaquette operator. Furthermore, the doublet $h_y+h_z$ receives a
logarithmic one-loop mass term from the top Yukawa coupling. These mass
contributions are parametrically of the order of the electroweak scale $v\sim
f/(16 \pi^2)$. The \tp-even doublet $h_y+h_z$ will become the dominant component
of the light Higgs boson, which is responsible for electroweak symmetry breaking.

Including all the aforementioned contributions, the scalar mass terms are given
by
\begin{align}
{\cal L}_{\rm mass,scal} =
-\frac{1}{2}\bigl [ & (M_p^2+m_{g,S}^2+m_{p,S}^2) \eta_x^2 +
    (M_t^2+m_{g,S}^2+m_{p,S}^2)(\eta_y^2 + \eta_z^2) + M_y^2 (\eta_y-\eta_z)^2
    \bigr ] \nonumber \\
-\frac{1}{2}\bigl [ & (M_p^2+m_{g,T}^2+m_{p,T}^2) |\phi_x|^2 + 
    (m_{g,T}^2+m_{p,T}^2) (|\phi_y|^2 + |\phi_z|^2) + 
    \tfrac{3}{2} M_y^2 |\phi_y-\phi_z|^2 \bigr ] \nonumber \\
-\frac{1}{2}\bigl [ & M_p^2 |h_x|^2 +
    (M_t^2+ m_{g,D}^2+m_{p,D}^2) (|h_x|^2 + |h_w|^2) +
    (M_{kin}^2 + M_y^2)|h_w-h_x|^2 \nonumber \\
    &+ (M_{kin}^2 + M_y^2)|h_y-h_z|^2 + m_t^2 |h_y+h_z|^2 + 
    (m_{g,D}^2+m_{p,D}^2) (|h_y|^2+|h_z|^2) \nonumber \\
    &+ i \, m_{\epsilon}^2 (h_w^\dagger h_x - h_x^\dagger h_w +
    	h_z^\dagger h_y - h_y^\dagger h_z) \bigr ] , \label{Mscal}
\end{align}
where the singlet, triplet, and doublet mass terms are shown in the first,
second, and remaining lines, respectively.
The mass parameters are summarized in the following list:
\begin{align}
     M_{p}^2 &\equiv 4 \kappa f^2 && \text{plaquette mass} \nonumber \\
     m_{\epsilon}^2 &\equiv \tfrac{\sqrt{3}}{2} f^2 \text{Im}(\epsilon) &&
       \epsilon\text{-plaquette term from \eqref{epsplaq}} \nonumber \\
     M_{\rm kin}^2 &\equiv c_k g^2 f^2 && \text{2-loop mass from \eqref{nonlin}} \nonumber \\
     M_{y}^2 &\equiv c_y \lambda_c^2 f^2 && \text{1-loop mass from \eqref{yuk1}}\nonumber \\
     m_{g,X}^2 &\equiv c_{g,X} \, g^4 f^2 / (4\pi)^2 \log(g^2 f^2/\Lambda^2) &&\text{gauge-loop mass, log part} \nonumber \\
     M_{t}^2 &\equiv c_T \lambda^2 f^2 && \text{top loop from \eqref{topyuk}, quadratic divergent part}\nonumber \\
     m_{t,D}^2 &\equiv c_t  M_{T'}^2/(4 \pi)^2 \log(M_{T'}^2/m_t^2) && \text{top loop from \eqref{topyuk}, log part} \nonumber \\
     m_{p,X}^2 &\equiv c_{p,X} \, \kappa^2 f^2 / (4\pi)^2 \log(\kappa f^2/\Lambda^2)  &&\text{plaquette-loop mass, log part} \nonumber 
\end{align}
Here the ${\cal O}(f)$ terms are written in capital letters, while lower case is
used for the lighter mass terms. 
$m_t$ and $M_{T'}$ denote the top quark mass and the mass of the $T'$ quark.
The latter cancels the quadratic divergences in the top loop contribution to
the Higgs mass. The $c_i$ are ${\cal O}(1)$ coefficients, which, except for
$c_t$, depend on unknown details of the UV completion. However,
it is possible to determine the \emph{relative} contributions of the gauge loops
to the singlets, doublets, and triplets, which are given by $c_{g,S}=0$ (since
the singlets commute with all gauge generators), $c_{g,T}\sim 1/8$, and
$c_{g,D}\sim \frac{3}{64} [1+(g'/g)^4]$.

The doublet $h_w$ does not get eaten and remains in the physical spectrum. It
mixes with the other \tp-odd doublet $h_x$ to form two mass eigenstates $h_{\HH1}$
and $h_{\HH2}$ with ${\cal O}(f)$ masses.
$\lambda_c$ can be relatively large, leading to a
rather large splitting between the two masses, and to a large mixing. In the
limit of large $\lambda_c$, the \tp-odd doublet masses are approximately given
by $M_{\HH1}^2 \approx M_t^2 + M_p^2/2$ and
$M_{\HH2}^2 \approx M_{\HH1}^2+2M_y^2+2M_{\rm kin}^2$.


\subsection{Electroweak symmetry breaking}\label{sec:ewsb}

The plaquette interactions \eqref{plaquette} generate quartic couplings for the
\tp-even scalars, which can be written as
\begin{equation}
-\kappa \, \tr [y,z]^2.
\end{equation}
Additional quartic interactions emerge from the second plaquette term
\eqref{epsplaq} and from loop corrections but will be neglected at this point.

To further analyse the Higgs potential, it is useful to switch back to the basis (\ref{higgsb}) using
\begin{align}
    h_a &= \tfrac{1}{\sqrt{2}} (h_y + h_z),  &
    h_b &= \tfrac{1}{\sqrt{2}} (h_y - h_z).
\end{align}
In this basis, the quartic potential for the \tp-even doublets reads
\begin{align}
        V_4 = \frac{\kappa}{8} \left[ (h_a^\dagger h_a)(h_b^\dagger h_b)
		+ (h_a^\dagger h_b)(h_b^\dagger h_a)
		- (h_a^\dagger h_b)^2 - (h_b^\dagger h_a)^2 \right], 
	\label{higgsquart}
\end{align}
while the quadratic potential, taken from eq.~\eqref{Mscal}, is given by
\begin{align}
    V_2 = \frac{1}{2}\bigl [ 
    m_a^2 | h_a|^2 + m_b^2 |h_b|^2 + 
    (m_{ab}^2 h_a^\dagger h_b + \mathrm{h.c.}) \bigr ]\,,
    \label{higgsquad}
\end{align}
with the mass parameters
\begin{align}
    m_a^2 &= 2m_t^2 +m_{g,D}^2 +m_{p,D}^2 \,,\\
    m_b^2 &= 2M_{kin}^2 + 2M_y^2 +m_{g,D}^2 +m_{p,D}^2\,, \\
    m_{ab}^2 &= -i m_\eps^2 \,.
\end{align}
As evident from these equations, electroweak symmetry breaking is described in
this model by an effective Two-Higgs-Doublet model (2HDM). 
The conditions for successful symmetry breaking are
\begin{align}
    m_{g,D}^2 &> -2 m_t^2\,, &
    m_\epsilon^4 &> (2 M_y^2 + 2 M_{kin}^2 + m_{g,D}^2 + m_{p,D}^2)
     (2 m_t^2 + m_{g,D}^2 + m_{p,D}^2)\,.
\end{align}
Since $m_{g,D}^2$ and $m_t^2$ are of the same order of magnitude and $m_t^2$ is
negative, these conditions can be satisfied naturally. 
Without the $m_\epsilon$ term, the Higgs potential would have an unstabilized flat
direction, and electroweak symmetry would not be broken to the SM vacuum.

The potential is then minimized by the vacuum expectation values
\begin{align}
    \langle h_a \rangle &= (0, v\cos \beta)^\top &
    \langle h_b \rangle &= (0, i \,v\sin \beta)^\top\,,
    \label{vevs}
\end{align}
with
\begin{equation}
\tan^2\beta = m_a^2/m_b^2 = {\cal O}(m^2/M^2),
\label{mixbet}
\end{equation}
where $M$ denotes the ${\cal O}(f)$ masses in the scalar potential, while $m$
represents any of the suppressed mass terms.
We have checked numerically that for reasonable choices of the mass parameters
defined above a value for $v$ close to the electroweak scale $v = 246 \gev$ can
be obtained.

The complex coupling constant $\eps$ of the second plaquette term
\eqref{epsplaq} leads to CP violation in the Higgs sector, as evident by the
complex vacuum expectation value of the second Higgs doublet in
eq.~\eqref{vevs}. Since it is assumed that $|\epsilon|$ is smaller than
$|\kappa|$ by about one order of magnitude, the amount of CP violation is
relatively small. Nevertheless, it could lead to potentially important
consequences for flavor physics. However, a detailed analysis of
CP-violating effects of our model is beyond the scope of this article and is
left for future work.

Neglecting the CP-violating contribution from $\epsilon$ and $m_\eps^2$, the
decomposition of the Higgs doublets into physical states is given by
\begin{align}
h_a &=
\begin{pmatrix}
\sqrt{2} G^+ \\ v + h^0 + i G^0
\end{pmatrix}
&
h_b
&= \begin{pmatrix}
\sqrt{2} H^+ \\ H^0 + i A^0
\end{pmatrix}.
\end{align}
As usual for a 2HDM,
one obtains the Goldstone bosons $G^0$, $G^+$ and $G^-=(G^+)^\dagger$, which
are eaten by the SM gauge bosons, a neutral pseudoscalar $A^0$, a pair
of charged scalars $H^+$ and $H^- = (H^+)^\dagger$, and two CP-even neutral
scalars $h^0$ and $H^0$.
The pseudoscalar mass is given by $M_A^2 = 
(m_a^2+m_b^2)$. The masses of $H^\pm$ and $H^0$ are very close to $M_A$,
differing only by ${\cal O}(m^2/M^2)$ effects. 

Including the CP-violating contribution from the $m_\eps^2$ parameter would lead
to a small mixing between the doublets and between CP eigenstates. However, as
mentioned above, these effects will be neglected for the purpose of this work.

The SM-like Higgs boson is $h^0$, which at tree-level has a very small mass, in
conflict with direct search limits\footnote{The small tree-level value for $m_h$
is not an artifact of our implementation of \tp-parity, but would also arise in
earlier versions of the minimal moose model in
Refs.~\cite{minimalmoose,Kilic:2003mq,LHT}.}. However, loop corrections to the
quartic potential yield positive contributions to $m_h$. For example, loops
involving the top quark and its heavy partners generate a correction of the type
\begin{equation}
\Delta m_h^2 \propto \frac{1}{\pi^2} v^2 \lambda_t^4 .
\end{equation}
In general, these radiative corrections cannot be computed explicitly in the
effective little Higgs theory, since they depend on the UV cutoff $\Lambda$.
However, they are generally comparable to the electroweak scale and thus could
lead to a value of $m_h$ above the current search limit.
Since $m_h$ is very sensitive to these loop contributions, we will take it as a
free parameters in the following.
Note that the loop corrections to the quartic potential have a negligible
effect on the masses of the heavy Higgs bosons $A^0, H^\pm, H^0$.


\section{Phenomenology}
\label{sc:pheno}

In Table~\ref{partlist} the particle content of the model beyond the SM gauge
bosons and fermions is summarized. Since the model requires a UV completion,
additional degrees of freedom are expected at the scale $\Lambda \sim 10$~TeV,
but will not be discussed here.
\begin{table}[tb]
\begin{center}
\renewcommand{\arraystretch}{1.2}
\begin{tabular}{|lc|c|c|c|}
\hline
    Field & & \tp-parity & \tpp-parity & Mass squared \\ 
\hline\hline 
Heavy gauge bosons & $B_{\HH\mu}^0$ & $-$ & $-$ & $\frac{4}{3}g'^2f^2$ \\
 & $W_{\HH\mu}^0, W_{\HH\mu}^\pm $ & $-$ & $-$ & $4g^2f^2$ \\
\hline
Singlet scalars & $\eta_x$ & $-$ & $-$ & $4\kappa f^2$ \\
 & $\eta_b \equiv \frac{1}{\sqrt{2}}(\eta_y-\eta_z)$ & 
 	$+$ & $-$ & $(2c_y\lambda_c^2+c_T\lambda^2)f^2$ \\
 & $\eta_a \equiv \frac{1}{\sqrt{2}}(\eta_y+\eta_z)$ & 
 	$+$ & $-$ & $c_T\lambda^2f^2$ \\
\hline
Triplet scalars & $\phi_x$ & $-$ & $-$ & $4 \kappa f^2$ \\
 & $\phi_b \equiv \frac{1}{\sqrt{2}}(\phi_y-\phi_z)$ & 
 	$+$ & $-$ & $3 c_y\lambda_c^2f^2$\\
 & $\phi_a \equiv \frac{1}{\sqrt{2}}(\phi_y+\phi_z)$ & 
 	$+$ & $-$ & $m_{g,T}^2 +m_{p,T}^2$ \\
\hline
\tp-odd doublet scalars & $h_{\HH1}$ & $-$ & $+$ & $M_{\HH1}^2$ \\
 & $h_{\HH2}$ & $-$ & $+$ & $M_{\HH2}^2$ \\
\hline
\tp-even doublet scalars & $H^\pm$ & $+$ & $+$ & $M_A^2$ \\
 & $A^0$ & $+$ & $+$ & $M_A^2$ \\
 & $H^0$ & $+$ & $+$ & $M_A^2$ \\
 & $h^0$ & $+$ & $+$ & $m_h^2$ \\
\hline
Heavy top partners & $T_\HH$ & $-$ & $-$ & $4\lambda_c^2 f^2$ \\
 & $T'$ & $-$ & $-$ & $4\lambda^2 f^2$ \\
 & $T$ & $+$ & $+$ & $4(\lambda^2+\lambda_c^2) f^2$ \\
\hline
Other heavy quarks & $Q_\HH$ & $-$ & $-$ & $4\lambda_c^2f^2$ \\
\hline
Heavy leptons & $L_\HH$ & $-$ & $-$ & $4(\lambda_c^l)^2f^2$ \\
\hline
\end{tabular}
\end{center}
\vspace{-2ex}
\mycaption{List of particles (besides SM particles) below the strong scale
$\Lambda$ and the dominant contributions to their masses. 
Mass corrections of order ${\cal O}(v/f)$ are neglected.\label{partlist}}
\end{table}

The charge eigenstates of the gauge bosons and scalars are given by
\begin{align}
W_\HH^0 &\equiv W_\HH^3, &
W_\HH^\pm &\equiv (W_\HH^1 \mp i W_\HH^2)/\sqrt{2}, \\[1ex]
\phi_i^0 &\equiv \phi_i^3, &
\phi_i^\pm &\equiv (\phi_i^1 \mp i \phi_i^2)/\sqrt{2}.
\end{align}

Most new particles have ${\cal O}(f) \sim {\cal O}$(TeV) masses. In the table,
relative corrections of order ${\cal O}(v/f)$ to these mass parameters have been
neglected. However, besides the light Higgs boson $h^0$, an additional scalar
triplet $\phi_a$ with weak-scale mass is predicted. These scalars are odd under
\tpp-parity, so that sizable numbers can be produced only in pairs, but since
they are even under \tp-parity, they can decay through the WZW coupling. In
principle, the WZW term also permits single $\phi_a$ production, but at a highly
suppressed rate, which is thus completely negligible. For the same reason, all
other \tpp-odd particles will decay first to one of the particles in $\phi_a$
through \tpp-conserving channels instead of directly decaying via the WZW term.

Since \tp-parity is exactly preserved, the lightest \tp-odd particle is stable.
If all coupling parameters are not much smaller than unity, the lightest \tp-odd
particle is the heavy U(1) gauge boson, $B^0_{\HH\mu}$, which is a viable dark
matter candidate.


\subsection{Electroweak precision constraints}

\tp-parity has been shown to largely reduce the constraints on the parameter
space in the case of the littlest Higgs model \cite{LHT,Hubisz:2005tx}, since
corrections to the electroweak precision observables arise only at loop level. 
Here we
calculate the corrections to the electroweak $S$ and $T$ parameters
\cite{Peskin:1991sw} in our model to determine the allowed parameter
space. 

The dominant contribution to $S$ and $T$ from the fermion sector come from gauge
boson self energy diagrams with the \tp-even $T$ quark running in the loop, a
contribution that has already been calculated in Ref.~\cite{Lavoura:1992np}.  
In
spite of the different symmetry structure of the model and the modified
implementation of the top-Yukawa couplings the results are almost identical to
those obtained in the case of the littlest Higgs model \cite{Hubisz:2005tx}. 
We find
\begin{align}
    \Delta S & = \frac{s_L^2}{2 \pi} \left[ c_L^2 \left( \frac{(m_T^2+m_t^2)^2}{(m_T^2-m_t^2)^2} - \frac{8}{3}\right) + \left( \frac{1}{3} + c_L^2 \frac{2 m_t^4 m_T^4(m_t^2-3m_T^2)}{(m_t^2-m_T^2)^3} - c_L^2  \right) \log \frac{m_t^2}{m_T^2} \right], \\
    \Delta T & = \frac{3}{16\pi} \frac{s_L^2}{\cw^2 \sw^2} \frac{m_t^2}{m_Z^2}\left[ s_L^2 \frac{m_T^2}{m_t^2} -1 - c_L^2 - \frac{2 c_L^2}{1-x_t} \log \frac{m_t^2}{m_T^2} \right],
\end{align}
where $s_L$, $c_L$ are the mixing angles defined in (\ref{tmixL}) and $\sw,\cw$
are the sine and cosine of the Weinberg angle, respectively.
Inserting the leading order expressions for the mixing angles (\ref{sintexp})
and expanding the expressions in the limit  $m_t^2 \ll m_T^2$  one arrives at 
\begin{align}
    \Delta S & =\frac{1}{2 \pi} \frac{\lambda^2}{\lambda_c^2} \frac{m_t^2}{m_T^2}\left( -\frac{5}{3} + \frac{2}{3}\log \frac{m_T^2}{m_t^2 }\right), \\
    \Delta T & =  \frac{3}{16 \pi}\frac{1}{\sw^2 \cw^2}\frac{\lambda^2}{\lambda_c^2}\frac{m_t^4}{m_T^2m_Z^2} \left( 2 \log \frac{m_T^2}{m_t^2} - 2 + \frac{\lambda^2}{\lambda_c^2} \right).
\end{align}
Another contribution to the $T$ parameter arises from the custodial symmetry
violating mass splitting  between the neutral and the charged $W_\HH$ gauge
bosons. At the one loop level this yields \cite{LHT,Hubisz:2005tx}:
\begin{align}
    \Delta T_{W_\HH} & = -\frac{9}{16 \pi \cw^2 \sw^2 M_Z^2} \Delta M^2_{W_\HH} \log\frac{\Lambda^2}{M^2_{W_\HH}}.
\end{align}
The logarithmic divergence forces one to introduce an appropriate counterterm
with an unknown coefficient $\delta_c$ of order one \cite{LHT,Hubisz:2005tx}. In our
model the mass splitting is given by
\begin{align}
    \Delta M^2_{W_\HH} & = \frac{g^2}{16}\frac{v^4}{f^2}\left(3+\sin^2 (2\beta)-\cos^2(2\beta)\right) \approx \frac{g^2}{8} \frac{v^4}{f^2} 
\end{align}
Including the counterterm this leads to a contribution to the $T$ parameter of
\begin{align}
    \Delta T_{W_\HH} & = -\frac{1}{4 \pi \sw^2} \frac{v^2}{f^2} \left( \delta_c + \frac{9}{4} \log \frac{4 \pi}{g} \right).
\end{align}
Next we discuss the contributions to electroweak precision observables that
arise from the scalar sector. The scalar singlets present in the theory do not
contribute to the $S$ and $T$ parameter. Contributions of the scalar triplets to
the $T$ parameter are proportional to the mass splitting between the charged and
neutral components. This splitting is induced only after electroweak symmetry
breaking and is generally small in our model, even for the light \tp-even
triplet.

In the limit of vanishing CP violation in the Higgs sector the contribution of
the two \tp-even Higgs doublets is well approximated by the SM Higgs
contribution and the contribution of a heavy Higgs doublet that is given by 
\cite{higgshunters}
\begin{align}
    \Delta T_{\rm 2HDM} & = \frac{1}{16 \pi \sw^2 \cw^2 m_Z^2} \left[ F (M_{H^+}^2, M_{A^0}^2) +  F(M_{H^+}^2,M_{H^0}^2) -F(M_{A^0}^2,M_{H^0}^2) )\right], \label{T2hdm}
\end{align}
where 
\begin{align}
    F(m_1^2,m_2^2) &= \frac{1}{2}(m_1^2 + m_2^2 ) - \frac{m_1^2 m_2^2}{m_1^2-m_2^2}\log \frac{m_1^2}{m_2^2} \,. 
\end{align}
For small mass differences this contribution is proportional to the mass differences between the charged and neutral heavy Higgs bosons. 
Since the actual values of $M_{H^\pm}^2 - M_{A^0}^2$ and $M_{H^0}^2 - M_{A^0}^2$  depend on unknown counterterm coefficients and are furthermore sensitive to radiative corrections to the quartic couplings, we take these mass differences as free parameters $\delta_\pm^2$ and $\delta_0^2$ of order $(100 \gev)^2$. 
The contribution to the $S$ parameter is small when the mass differences of the heavy scalars are small compared to their masses, so we can neglect it here. 
Taking into account the CP violation in the Higgs sector only affects the mixing between the Higgs scalars. Since these mixings are small, they do not change the contributions to the $S$ and $T$ parameter significantly.  

The \tp-odd doublets lead to a contribution of similar size, which depends on
the incalculable ${\cal O}(v^2)$ mass splittings in the $h_w$ and $h_x$
doublets. For simplicity, we do not include these terms explicitly, since
the overall magnitude of the Higgs corrections can be estimated sufficiently
well from equation (\ref{T2hdm}).

Other one loop contributions to the $T$ parameter arise from mass splittings in the
mirror fermion doublets. The magnitude of such corrections has been estimated in
Ref.~\cite{Hubisz:2005tx} and it was found that they are suppressed compared to the
contributions discussed above. 

Apart from the loop-induced contributions to the $T$ parameter the custodial
symmetry violating kinetic term of the Goldstone bosons \eqref{LG} contributes at
the tree level through operators of the form\footnote{We thank Ian Low for
pointing out the relevance of this operator to us.}
\begin{align}
  \frac{c}{f^2} \left| h_{a,b}^\dagger D_\mu h_{a,b}\right|^2\,,
\end{align}
where $h_{a,b}$ are the \tp-even Higgs doublets.
In our model this leads to a sizable contribution to the $T$ parameter of 
\begin{align}
  \Delta T \approx 0.5 \frac{\tev}{f^2}\,. \label{custT}
\end{align}
This contribution seems to disfavor values of $f$ below about $2 \tev$. 
However, it will be shown below that values of $f$ around $1 \tev$ can be in
agreement with experimental data due to cancellations between different
contributions to the $T$ parameter\footnote{Note that a larger custodial
symmetry violating contribution from heavy gauge boson exchange in the model of
Ref.~\cite{Kilic:2003mq} is forbidden by \tppp-parity.}. 

The experimental values for the $S$ and $T$ parameters are \cite{Amsler:2008zzb}
\begin{align}
    S & = -0.04 \pm 0.09 \\
    T & = 0.02 \pm 0.09
\end{align}
for a Higgs mass of $m_h = 117\gev$ and fixing the $U$ parameter to $U=0$. The
contributions to the $S$ parameter from the top sector are small for all
reasonable choices of parameters, and in particular do not lead to additional
constraints on regions that give a satisfactory $T$ parameter.

\begin{figure}[tb]
\begin{minipage}[b]{7.9cm}
\psfig{figure=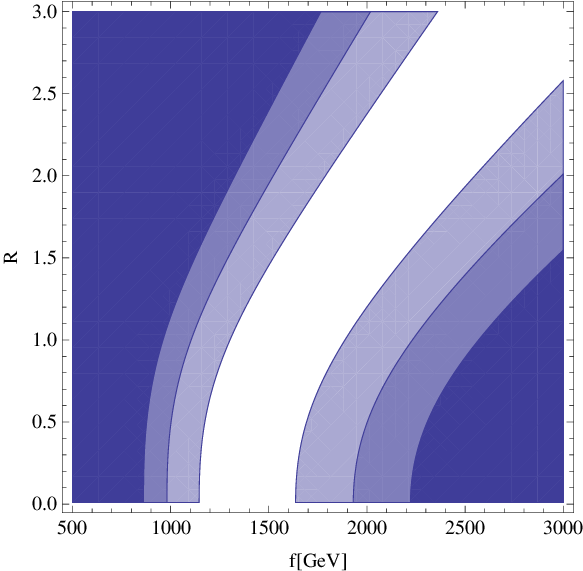, width=7.8cm}
\end{minipage}%
\hfill%
\begin{minipage}[b]{8cm}
\psfig{figure=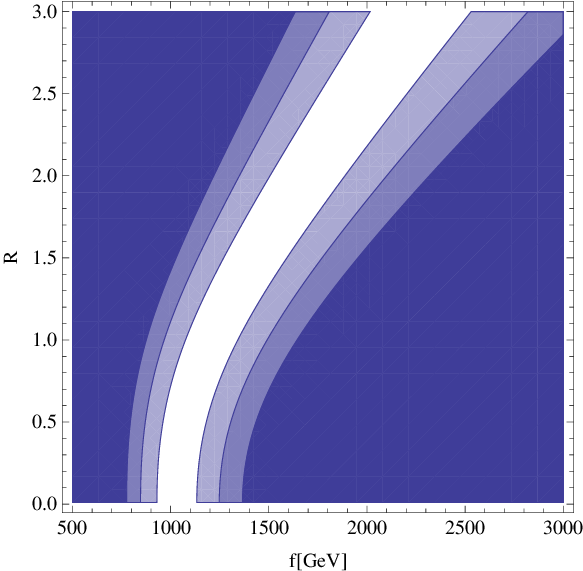, width=7.8cm}
\end{minipage}
\mycaption{Allowed regions in the $f$-$R$ parameter space for fixed values of
$\delta_c$, $\delta_\pm$ and $\delta_0$. From lightest to darkest the shaded
regions indicate a deviation of the $T$ parameter from the experimental value by
more than one, two and three sigma, respectively. 
Both plots use $\delta_c=5$. The mass splittings are $\delta_\pm^2 = 0.1 f^2$ and $\delta_0^2 = 0.2 f^2$ in the left plot and $\delta_\pm^2 = -0.15 f^2$ and $\delta_0^2 = -0.3 f^2$ in the right plot.
\label{f-Rplot}}
\end{figure}

For values of $f>1 \tev$ the contributions of the top sector and the gauge boson
sector each stay within the experimental limit of $T$ for most choices of
the parameters $R$ and $\delta_c$ respectively. The contribution
\eqref{custT}, taken separately, would push this value towards $f\gesim 2 \tev$. The
contribution of the Higgs doublets does not directly constrain the scale $f$ but
essentially depends on the  mass splitting $\delta_\pm$ and $\delta_0$. 
When the mass splittings are such that one neutral Higgs is
lighter and one heavier than the charged Higgs boson, this contribution is
negative and can partially cancel the
contribution \eqref{custT}, thus allowing lower values of $f$. In figure
\ref{f-Rplot} we show that for reasonable choices of the  mass splitting
parameters and of $\delta_c$ these cancellations take place, allowing for values
of $f$ at the $1 \tev$ scale, and even slightly below. 

The left plot shows an example where the $H^0$ is the heaviest Higgs boson and 
$H^\pm$ is heavier than $A^0$, while the right plot shows an example with the
hierarchy inverted. For both plots the splittings have been chosen proportional
to the mass scale $f$.  This causes some regions in the $f$-$R$ plane to be
excluded also for large values of $f$, since there the contributions from the
Higgs loops become dominant. 

A moderate amount of fine tuning is involved to cancel the contribution of
eq.~\eqref{custT} for smaller values of $f$. At this point it is worth
mentioning that this contribution  is absent in models where the Higgs sector
has a custodial symmetry, which can be achieved by enlarging the global symmetry
group. A concrete realization of this idea, based on a $SO(5)\times SO(5)$ group
structure, has been constructed {\it e.$\,$g.} in Ref.~\cite{Chang:2003un}. It
is certainly possible to extend the present model in a similar way in order to
enlarge the allowed parameter space at low scales, however for the sake of
simplicity we decided against discussing this here. Furthermore, while this
model allows a straightforward ultraviolet completion with QCD-like dynamics,
such a construction is less obvious for models that implement a custodial
symmetry using orthogonal groups.


\subsection{Decays of heavy particles}

\begin{figure}
\begin{minipage}[b]{7.9cm}
\fbox{\psfig{figure=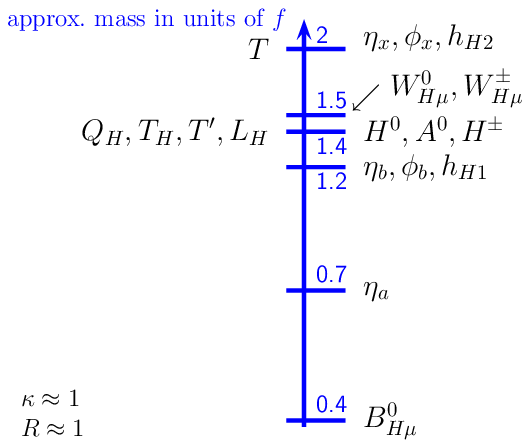, width=7.8cm, bb=170 465 460 680}}
\end{minipage}%
\hfill%
\begin{minipage}[b]{8cm}
\fbox{\psfig{figure=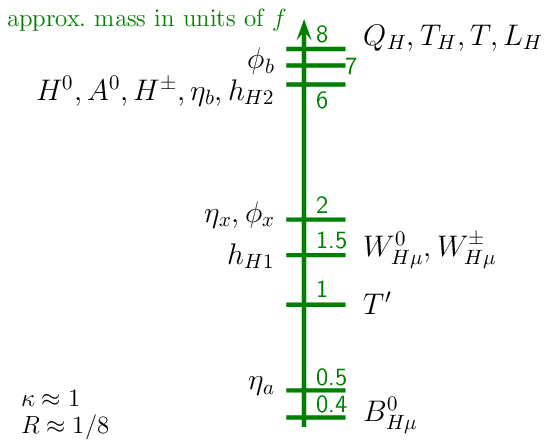, width=7.8cm, bb=170 465 460 680}}
\end{minipage}
\mycaption{Approximate patterns of two typical spectra of ${\cal O}(f)$ particle 
masses. In both cases, the non-calculable coefficients are assumed to be of
order unity, $c_i\approx 1$.
\label{hierarchy}}
\end{figure}

For concreteness, we assume the plaquette parameter to be close to unity,
$\kappa \approx 1$. Furthermore, the UV-sensitive coefficients $c_i$, introduced
below eq.~\eqref{Mscal}, are also assumed to be of order ${\cal O}(1)$. As
pointed out above, the Yukawa coupling $\lambda_c$ of the mirror fermions can be
chosen relatively large, $\lambda_c \gg 1$, since these fermions do not play any
role in compensating the quadratic divergences in the Higgs mass. In this case
also the \tp-even top partner $T$ will be heavy. As examples, two scenarios will
be considered, one with mirror fermion masses near the breaking scale $f$, and one
with very heavy mirror quarks:
\begin{align}
&\text{``Light mirror fermion'' scenario:} &&
\lambda_c \approx \lambda \approx 1/\sqrt{2}, && R \approx 1, \\[1ex]
&\text{``Heavy mirror fermion'' scenario:} &&
\lambda_c \approx 4,\;\; \lambda \approx 1/2, && R \approx 1/8.
\end{align}
Note that $\lambda_c$, $\lambda$ and $R=\lambda/\lambda_c$ are related through
the top Yukawa coupling
\eqref{tyuk}, which must be $\lambda_t \approx 1/\sqrt{2}$ to reproduce the
experimental value for the top-quark mass.
The
mass hierarchy of the two scenarios is sketched in
Fig.~\ref{hierarchy}.

The mass pattern and the conservation of \tp- and \tpp-parity and gauge
symmetries strongly 
constrain the possible decay channels of the heavy particles. The gauge
symmetries, however, are violated by electroweak symmetry breaking, leading to a
small mixing between the heavy
gauge bosons $W_\HH^{0}$ and $B_\HH^{0}$, with the mixing angle given by
\begin{equation}
\sin\theta_\HH = \frac{3gg'}{16(3g^2-g'^2)} \, \frac{v^2}{f^2}.
\end{equation}
While this mixing is suppressed by two powers of $v/f$, it nevertheless 
can be relevant for decays of some particles that do not have any other possible
decay modes.

The dominant decay channels are summarized in Table~\ref{dec}, for the two
scenarios introduced above. Not included in the table are weakly interacting
particles with masses larger than about $2f$ and strongly interacting particles
with masses larger than about $5f$, since they are expected to be beyond the
reach of the LHC (assuming $f \gesim 500$~GeV). As mentioned above, the lightest
\tpp-odd particle decays through the WZW term, but the WZW contribution is
negligible compared to \tpp-conserving interactions for all decays of heavier
\tpp-odd particles.

Independent of other parameters, the lightest \tpp-odd particle will be one of
the scalars in the triplet $\phi_a$, since they do not receive any ${\cal O}(f)$
mass terms. At leading order in $1/f$ the WZW term induces decays of into 
pairs of SM gauge bosons \cite{WZW}.
The masses of the three scalars $\phi_a^{0,\pm}$ are almost
degenerate, with a small splitting between the neutral $\phi^0_a$ and the
charged $\phi^\pm_a$ incurred from EWSB and 
the gauge boson loop contribution $m_{g,T}$ in eq.~\eqref{Mscal} only at order 
${\cal O}[g^4f^2/(4\pi^2)] \sim {\cal O}[g^4v^2]$. At this order,
higher-order operators from the UV completion could yield additional
contributions to the mass splitting, so that it cannot be calculated reliably
from the effective little Higgs model. For concreteness, we will therefore
assume that the $\phi^\pm_a$ are slightly heavier than $\phi^0_a$, opening up
the decay $\phi^\pm_a \to (W^\pm)^* \phi^0_a$  through a virtual $W$ boson. 
Depending on the magnitude of the mass splitting, this decay could dominate
over the direct decays into $W^\pm \gamma$ and $W^\pm Z$ that are 
mediated by the WZW term. 
\begin{table}[tb]
\begin{center}
\renewcommand{\arraystretch}{1.2}
\begin{tabular}{|l|l|}
\hline
\makebox[5cm][l]{``Light mirror fermions''} & 
\makebox[5cm][l]{``Heavy mirror fermions''} \\
   $\;\;\kappa\approx 1, R\approx 1$ &
   $\;\;\kappa\approx 1, R\approx 0.09$ \\
\hline \hline
$Q_\HH \to q \,B_\HH^0;\qquad L_\HH \to l \,B_\HH^0$ & \\
$T' \to t \,B_\HH^0$ & $T' \to t \,B_\HH^0$ \\
$T \to t \, h^0, \;  t \, Z, \;  b \, W^+, \; t \, H^0, \; t \, A^0, \; b \, H^+, \; T' \, B^0_\HH$ & \\
\hline
$W^0_\HH \to 
  \bar{f} \, F_\HH, \; f\,\overline{F}_\HH$ %
 & $W^0_\HH \to h^0 \, B^0_\HH$ \\
$W^\pm_\HH \to 
  \bar{f}' \, F_\HH, \; f'\,\overline{F}_\HH$ %
 & $W^\pm_\HH \to W^\pm B^0_\HH$ \\
\hline
$H^0 \to t\bar{t}$ 
 & \\
$A^0 \to t\bar{t};\qquad H^+ \to t\bar{b}$ & \\
\hline
$h_{\HH 1}^{0,\pm} \to t\bar{t}\,\phi_a^{0,\pm}\,B_\HH^0$ &
$h_{\HH 1}^{0,\pm} \to t\,\overline{T}{}'\,\phi_a^{0,\pm},\;
			\bar{t}\,T'\,\phi_a^{0,\pm}$ \\
\hline
$\eta_a \to h^0 \, \phi^0_a$ & $\eta_a \to h^0 \, \phi^0_a$ \\
$\phi_a^0 \to Z^0Z^0$, $Z ^0\gamma$, $\gamma\gamma$ & $\phi_a^0 \to Z^0Z^0$, $Z^0 \gamma$, $\gamma\gamma$ \\
$\phi_a^\pm \to (W^\pm)^* \phi^0_a$, $W^\pm Z$, $W^\pm \gamma$ & $\phi_a^\pm \to (W^\pm)^* \phi^0_a$, $W^\pm Z$, $W^\pm \gamma$ \\
\hline
$\eta_b \to (A^{0})^*\,\phi^0_a,\; (H^{\pm})^* \, \phi^\mp_a,\;
  (A^{0})^*\,\eta_a$
  & \\
$\phi_b^0 \to  (A^{0})^*\,\phi^0_a,\; (H^{\pm})^* \, \phi^\mp_a,\;
  (A^{0})^*\,n_a$ & \\
$\phi_b^\pm \to h^0\,\phi^\pm_a$ & \\
\hline
\end{tabular}
\end{center}
\vspace{-2ex}
\mycaption{Dominant decay modes for heavy particles expected to be observable at
the LHC, for the two qualitative spectra in Fig.~\ref{hierarchy}. Weakly
interacting particles with masses $M\gesim 2f$ and strongly interacting particles
with masses $M\gesim 5f$ are not listed, since they are assumed to be beyond the
reach of the LHC. $(X)^*$ indicates an off-shell particle.\label{dec}}
\end{table}

In the ``light mirror fermion'' scenario,
since the SU(2) gauge bosons $W_\HH^{0,\pm}$ are relatively heavy, they can
decay into a mirror fermion plus the corresponding SM partner fermion.
Decays of $W_\HH^{0,\pm}$ directly to the lightest \tp-odd particle $B_\HH^0$
via emission of SM gauge bosons or Higgs bosons 
are suppressed by ${\cal O}(v^2/f^2)$. Therefore, these channels have a
branching ratio of at most a few per-cent.
Similarly, to leading order in
$v/f$,
the other two top partners $T$ and $T'$ are SU(2) singlets and thus only
interact through Yukawa or U(1) couplings. Consequently, the do not contribute
significantly to heavy SU(2) gauge boson decays.

The \tp-odd fermions can only decay to the heavy hypercharge boson $B_\HH^0$.
Although the mirror fermions are not charged under the heavy hypercharge group
(see Table~\ref{ferm}), this decay is enabled through the mixing between $W_\HH^{0}$
and $B_\HH^{0}$.

\begin{figure}
\begin{center}
\epsfig{figure=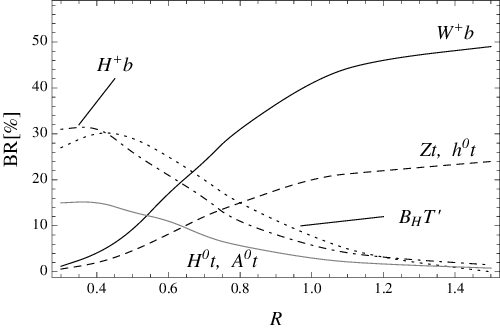, width=10cm}%
\end{center}
\vspace{-2em}
\mycaption{Branching fractions for the dominant decay modes 
of the \tp-even $T$ quark, as a function of $R=\lambda/\lambda_c$, and for
$f=1$~TeV. \label{Tbr}} 
\end{figure}

The situation is different for the \tp-even $T$ quark, which has sizable
couplings to the Higgs bosons from the top Yukawa term \eqref{topyukcooked}
and to the  $B_\HH^0$ boson via its hypercharge quantum number.
Figure~\ref{Tbr} shows the branching fractions for the dominant decay modes
as a function of the Yukawa coupling ratio $R$. For the purpose of this plot,
the Higgs boson masses have been calculated using the loop-induced mass terms
from section \ref{sc:scalars} with $c_i=1$.
The branching ratios depend only mildly on $f$. For $R>1$, the decays into $W^+b$, $Zt$ and $h^0t$ dominate, with relative branching fractions dictated by the Goldstone boson equivalence theorem. Decays into the heavyer Higgs bosons become sizable for smaller values of $R$, where in addition the parity odd mode $T\to B_H T'$ becomes relevant. 

In the second scenario, the mirror fermions and many scalar particles are too
heavy to be observables at the LHC. In this case, the gauge bosons $W_\HH^{0,\pm}$ decay
to the $B_\HH^0$ via emission of a SM gauge boson or the little Higgs boson. 
As mentioned above, the $T'$ top partner, which is always lighter than the heavy SU(2)
gauge bosons, is a SU(2) singlet. As a result, the decay $W_\HH^+ \to T' \bar{b}$
is forbidden, while the channel $W_\HH^{0} \to T' \bar{t},\overline{T}{}' t$ 
can only proceed through the small mixing of the $W_\HH^{0}$ with the $B_\HH^{0}$.
Therefore this leads to an additional suppression compared to the decay $W_\HH^{0}
\to h^0 \, B_\HH^{0}$:
\begin{equation}
\Gamma[W_\HH^{0} \to T' \bar{t},\overline{T}{}' t] \propto
\cos^2\theta_\HH \approx 10^{-3} \times v^4/f^4,
\qquad
\Gamma[W_\HH^{0} \to h^0 \, B_\HH^{0}]\ \propto v^2/f^2.
\end{equation}
Consequently, the decay of the heavy SU(2) gauge bosons into the top partner
$T'$ can be neglected.

The \tp-even Higgs bosons $H^0$, $A^0$, and $H^\pm$ decay predominantly into
third-generation SM fermions through the Yukawa couplings
eqs.~\eqref{yuk1},\eqref{topyuk}. On the other hand, their coupling to the SM
gauge bosons is suppressed by the small mixing angle $\beta$, see
eq.~\ref{mixbet}, rendering these decay channels negligible.

Of the \tp-odd doublet scalars, one doublet is typically very heavy. The lighter
doublet $h_{\HH 1}$ contains one CP-even and one CP-odd neutral scalar and two
charged states. Their decays are strongly constrained by their charges
under \tp- and \tpp-parity. If the $T'$ is light enough, three-body decay
channels are open, otherwise the scalars in  $h_{\HH 1}$ can only decay into a
four-body final state.

For the singlet scalars the plaquette
operator \eqref{plaquette} is the only interaction term in the model.
At tree-level, the $\eta_b$ singlet
can decay into $A^{0}\,\phi^0_a$, $H^{\pm} \, \phi^\mp_a$, and $A^{0}\,\eta_a$, 
which all have partial widths of the roughly the same order. 
As the masses
of $\eta_b$, $A^0$ and $H^\pm$ are close to each other, 
the doublet Higgs bosons must be slightly off-shell in these decays.
In the same way one obtains the decay modes of $\phi_b^{0,\pm}$.


\subsection{Collider phenomenology}

For values of $f$ near 1~TeV, several of the new particles predicted by the
minimal moose model with exact \tp-parity are within reach of the LHC.  We have
calculated cross sections using the program  {\sc CompHEP 4.4} \cite{comphep},
using a model file generated with the help of  the {\sc LanHEP}
package \cite{lanhep}.

The production of heavy gauge bosons ($W_\HH^{0,\pm}$) and mirror quarks
($Q_\HH$) proceeds in the same way as for the littlest Higgs model with
\tpp-parity, since all
relevant interactions are constrained by gauge invariance. The reader is
referred to the literature on the littlest Higgs model for more details on
production channels and cross sections \cite{lhtpheno}. 
However, compared to the littlest Higgs model, the \tp-odd gauge bosons are
heavier in the minimal moose model (as a function of $f$). As a result,
production cross section for these heavy gauge bosons are relatively small
throughout the allowed parameter range.

A special feature of our model are the light triplet scalars
$\phi_a^{0,\pm}$. Since they are odd under \tpp-parity, the single
production cross section is negligible, but pair production can lead to sizable
rates. The lightest \tpp-odd scalar, assumed to be the $\phi^0_a$, decays
through the WZW interaction into two SM gauge bosons. In particular, the decay
into two photons is allowed, leading to striking signatures with one charged lepton
or jet and up to four photons in the final state.

\begin{figure}
\begin{center}
\begin{minipage}[b]{5cm}
\psfig{figure=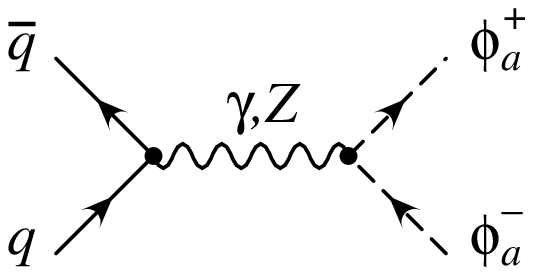, width=4.5cm}\\[1.5em]
\psfig{figure=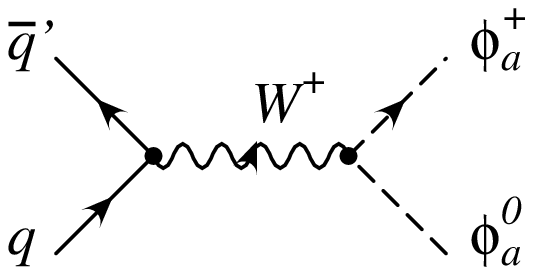, width=4.5cm}\\[1em]
\end{minipage}
\hspace{1em}
\epsfig{figure=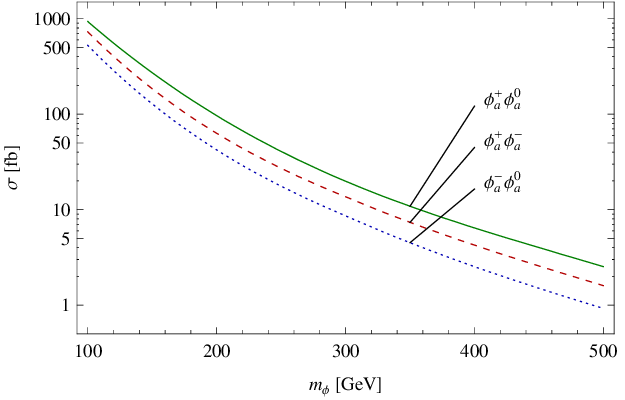, width=10cm}%
\end{center}
\vspace{-1.5em}
\mycaption{Pair production diagrams and
LHC cross sections for the particles in the lightest
scalar triplet, as a function of their mass. The factorization scale has been
set to $m_\phi$, and the center-of-mass energy is 
$\sqrt{s} =14\tev$.\label{phia}} 
\end{figure}

The main production mode for $\phi_a$ pairs at the LHC are the Drell-Yan
processes with the Feynman diagrams shown at the left of Fig.~\ref{phia}.
The tree-level cross sections are also shown in
Fig.~\ref{phia}.
The production of $\phi_a$ pairs from gluon fusion through s-channel Higgs boson
exchange is suppressed by several powers of $v/f$. We have checked explicitly
that this channel is negligible compared to the leading Drell-Yan mode. $W^\pm + 3 \gamma$
and $W^\pm + 4 \gamma$ are the most exciting final states that result from $\phi_a$ pair
production. 

For all other exotic scalars in the model the productions cross sections are
small, ${\cal O}$(fb) or below, since those particles are relatively heavy and
have only couplings of weak interaction strength. Therefore the observation of
any of these scalars from direct production at the LHC would be very
challenging.

On the other hand, colored particles have relatively large cross sections at the
LHC, in particular the top-quark partners $T$, which can be produced singly, and
$T'$, which is predicted to be relatively light to cancel the quadratic
divergences to the Higgs mass parameter.

\begin{figure} 
(a) \hfill \ \ \  (b) \hfill \ \\[-1em]
\epsfig{figure=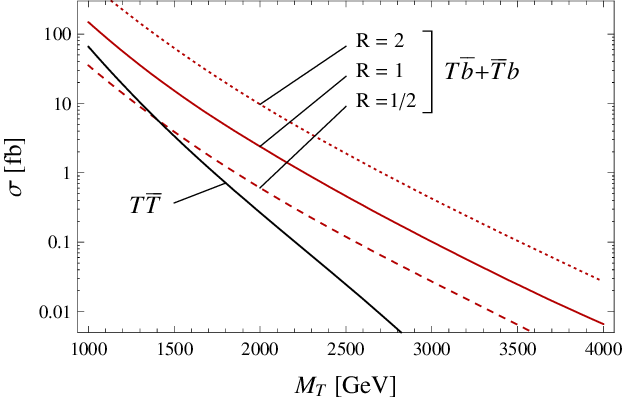, width=0.49\textwidth}\hfill
\epsfig{figure=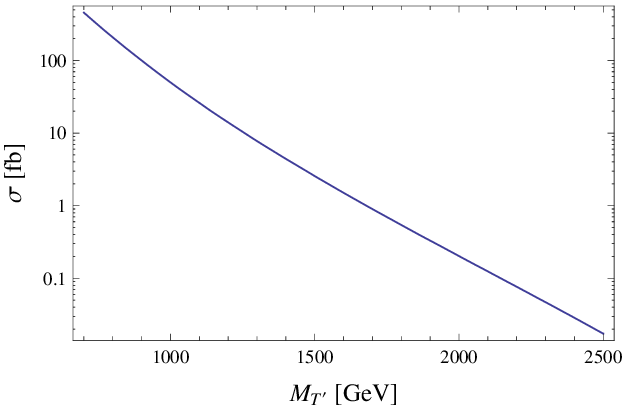, width=0.49\textwidth}%
\vspace{-1ex}
\mycaption{{\bf (a)} LHC cross sections for single $T$ and $T\overline{T}$ production, as a
function of the $T$ quark mass, and for different values of 
$R\equiv\lambda/\lambda_c$. {\bf (b)} LHC cross section for $T'\overline{T}{}'$
production, as a function of the $T'$ mass. In each plot,
the QCD and factorization scales have been
set to $M_{T^{(\prime)}}$, and the center-of-mass energy is 
$\sqrt{s} =14\tev$.\label{T}} 
\end{figure}

Single $T$ production, $pp \to T\bar{b}+X, \; \bar{T}b+X$ proceeds dominantly
through the partonic processes $b\bar{q} \to T \bar{q}'$ and $\bar{b}q \to
\overline{T} q'$, where $q,q'$ are SM quarks of the first two generations. The
initial-state bottom quarks can be thought of originating from gluon splitting,
$g \to b\bar{b}$, but for the purpose of this analysis we use the alternative 
formulation where the bottom quarks are included in the parton distribution
functions, see for example Ref.~\cite{bpdf}. $T$ quarks can also
be produced in pairs through the partonic processes $gg \to T\overline{T}$ and
$q\bar{q}\to T\overline{T}$. The LHC production cross sections are shown in
Fig.~\ref{T}~(a).

Single $T$ production is mediated mainly by t-channel exchange of $W$
bosons,  which  couple only to the small top-quark admixture in $T$, see
eq.~\eqref{sintexp}. As a result, the single $T$ cross section strongly depends
on the mixing parameters, and thus on $R=\lambda/\lambda_c$. In contrast, the
pair production process is mainly governed by QCD gluon exchange and thus
insensitive to mixing. In spite of the coupling suppression of the single $T$
contribution this process is dominant for $M_T \gesim 1$~TeV, owing to the
smaller mass of the final state system (see Fig.~\ref{T}~(a)).

For relatively low values of $M_T$, the production rates can reach several tens
of fb. For $R>1$ the dominant final states from single $T$ production are $Wbb$, $ZWbb$ and $4b+W$, where for the last mode we assume a light Higgs with dominant decay $h^0 \rightarrow b \bar{b}$. 
For smaller $R$ relevant final states include $3t+b$ and $ttbb$ from decays of the heavy Higgs bosons, and $tb+\Eslash_T$ from the parity odd decay mode.

Figure~\ref{T}~(b) shows the pair production cross section for $T'$ quarks.
Since it is quite possible that $M_{T'} < 1$~TeV, the cross section can
amount to several 100 fb. However, the decay $T'\overline{T}{}' \to
t\bar{t}\,B_\HH^0\,B_\HH^0$ leads to a signature that is very similar to
SM $t\bar{t}$ production and requires a careful analysis to disentangle from
this background \cite{lhtpheno,ttmiss}.

\vspace{1ex}
Besides new particle production, SM processes can be modified by the effect of
virtual heavy particle contributions. In particular, the production rate of the SM-like light Higgs boson
$h^0$ via gluon fusion can receive sizable corrections from loop diagrams
involving the heavy top partners. However, this effect is not unique to our
implementation of \tp-parity, but it is
completely analogous to the 
littlest Higgs model, described in detail in Ref.~\cite{Chen:2006cs}.


\section{Summary}
\label{sc:concl}

In this paper we present a little Higgs model where a new \tp-parity is
implemented such that it is not broken by operators that are typically
introduced in strongly coupled ultraviolet completions. This symmetry can
therefore be exact up to very high scales and in particular reestablishes the
lightest \tp-odd particle as a viable Dark Matter candidate for little Higgs
models. 

Our construction is based on the Minimal Moose little Higgs model. Following
\cite{Krohn:2008ye} we introduce \tp-parity as an exchange symmetry between
the link fields in the model. The gauge transformation properties of the link
fields are chosen such that the gauged WZW term is even under \tp-parity
while ensuring that the additional heavy gauge bosons present in the model
remain \tp-odd. An additional approximate $\mathbb{Z}_2$ symmetry further
restricts the interactions in the scalar sector and removes potentially
dangerous operators.  In the fermion sector a set of mirror-fermions is
introduced in order to implement \tp-parity without generating large four
fermion operators.  An additional pair of top quark partners is introduced to
avoid large breaking of the global symmetry that protects the Higgs mass. Mass
terms for the mirror fermions and the additional top quark are introduced in a
\tp-invariant way while preserving enough global symmetries to not
generate a large mass for the Higgs fields.

Below the symmetry breaking scale $f$, a light \tp-even Higgs boson and a scalar
triplet $\phi_a$ remains in the spectrum of the model. In addition, the masses
of the $B_H$ gauge boson and of the scalar singlet $\eta_a$ are parametrically
smaller than $f$. For all reasonable choices of parameters the $B_H$ is the
lightest \tp-odd particle and therefore the Dark Matter candidate, similar
to the original little Higgs models with \tp-parity. The Higgs sector has the
structure of a two-Higgs doublet model with one heavy doublet. Successful
electroweak symmetry breaking is achieved with moderate fine tuning of
parameters and yields a light physical Higgs boson. The model includes a number
of additional scalars, which do not acquire vacuum expectation values since they
are odd under one of the two parities. Most of these scalars obtain large ${\cal
O}(f)$ masses, except for the aforementioned $\phi_{a}$, which has a mass of
order the electroweak scale. 

The contributions to the electroweak $S$ and $T$ parameters from our model are
moderate, allowing for new physics scales as low as $f \sim 1 \tev$. 
This opens the possibility for the model
to be detectable at the LHC within the first years of running. In addition to
the usual decay signatures of little Higgs models, the light scalar triplet can
be pair-produced copiously at hadron colliders and yields a peculiar signature
from its main decay channels into photons or $W$ and $Z$ boson pairs.  Probing
the top quark sector of the model is more challenging since most signatures
suffer from a large standard model background. 
It would be interesting to study the phenomenological signatures of this model
in more detail, in particular whether it can be distinguished from other little
Higgs models. Also the question whether the $B_H$ can account for the observed
dark matter density in the universe remains to be answered. 

Our model is a realistic realization of the little Higgs mechanism 
with dark matter. More elaborate constructions can be envisaged 
where the parameter space is less constrained by low energy bounds.


\section*{Acknowledgements}

The authors would like to thank H.-C.~Cheng, A.~von Manteuffel 
and in particular I.~Low for useful discussions.
This project was supported in part by the Schweizer Nationalfonds.


\end{document}